# DesignCon 2007

## Power Supply Compensation for Capacitive Loads

Jonathan L. Fasig, Mayo Clinic

Barry K. Gilbert, Mayo Clinic
gilbert.barry@mayo.edu, 507-284-4056

Erik S. Daniel, Mayo Clinic

## Abstract
As ASIC supply voltages approach one volt, the source-impedance goals for power distribution networks are driven ever lower as well.  One approach to achieving these goals is to add decoupling capacitors of various values until the desired impedance profile is obtained.  An unintended consequence of this approach can be reduced power supply stability and even oscillation.  In this paper, we present a case study of a system design which encountered these problems and we describe how these problems were resolved.  Time-domain and frequency-domain analysis techniques are discussed and measured data is presented.

## Author(s) Biography
Jonathan L. Fasig received a BSEE from the Milwaukee School of Engineering (Milwaukee, WI).  He is currently a Principal Engineer at the Mayo Clinic Special Purpose Processor Development Group where he has been involved in signal integrity and power integrity.

Barry K. Gilbert received a BSEE from Purdue University (West Lafayette, IN) and a Ph.D. degree in physiology and biophysics from the University of Minnesota (Minneapolis, MN).  He is currently Director of the Special Purpose Processor Development Group at the Mayo Clinic, directing research efforts in high performance electronics and related areas.

Erik S. Daniel received a BA in physics and mathematics from Rice University (Houston, TX) in 1992 and a Ph.D. degree in solid state physics from the California Institute of Technology (Pasadena, CA) in 1997.  He is currently Deputy Director of the Special Purpose Processor Development Group at the Mayo Clinic, directing research efforts in high performance electronics and related areas.

## Introduction

Much has been written regarding strategies and methods to design power distribution networks that minimize disturbances on the system DC supply voltages, and it has become commonplace to invest considerable time and effort in the optimization of circuit-board constructions and decoupling-capacitor selections to ensure that the system power nets are disturbance-free [1,2,3]. So it can be a nasty surprise indeed when an oscilloscope is connected to a poorly-behaved prototype and significant "noise" is observed on a supply voltage. This paper describes the diagnosis of such a system and the steps we took to remedy it.

## System Overview

The focus of our interest was a system comprised of one large motherboard and two optical-interface daughter cards. This motherboard incorporated several very large field-programmable gate arrays (FPGA) along with various clock generation and support circuits. DC power for the system was supplied from a single AC-powered unit capable of generating multiple constant-voltage outputs between +12 V and -5.2 V.

Early in the design of the motherboard the designers became aware that significant DC current surges could be produced when the FPGAs transitioned from program-mode to functional-mode. To mitigate supply voltage droop during these transient events, the motherboard designers incorporated several large-capacitance low-ESR tantalum capacitors into the decoupling networks for the FPGA supplies.

In the course of prototype testing, it was discovered that the system exhibited various functional and performance limitations including difficulty programming the FPGA's and highly elevated bit error rates. Further investigation revealed that all these problems abated when the motherboard was powered from lab-grade bench supplies instead of the internal multiple-output power supply. No issues were observed when the outputs of the internal supply were monitored driving just purely resistive loads, but when the internal supply was connected to the motherboard then significant ringing was observed on the FPGA supplies coincident with steps in the load current as shown in Figure 1.

The salient difference between the purely resistive load (provided using an Agilent N3306A programmable load) and the system motherboard was the 4700 microfarads of bulk storage capacitance added to the motherboard power distribution networks. Clearly, our internal supply had compatibility issues with our motherboard bulk capacitors. If these capacitors were removed then the ringing abated as well, but the supply voltage at the FPGAs fell excessively during the current surge. Consequently, outright elimination of the bulk capacitors was not a viable solution. Another solution was required.

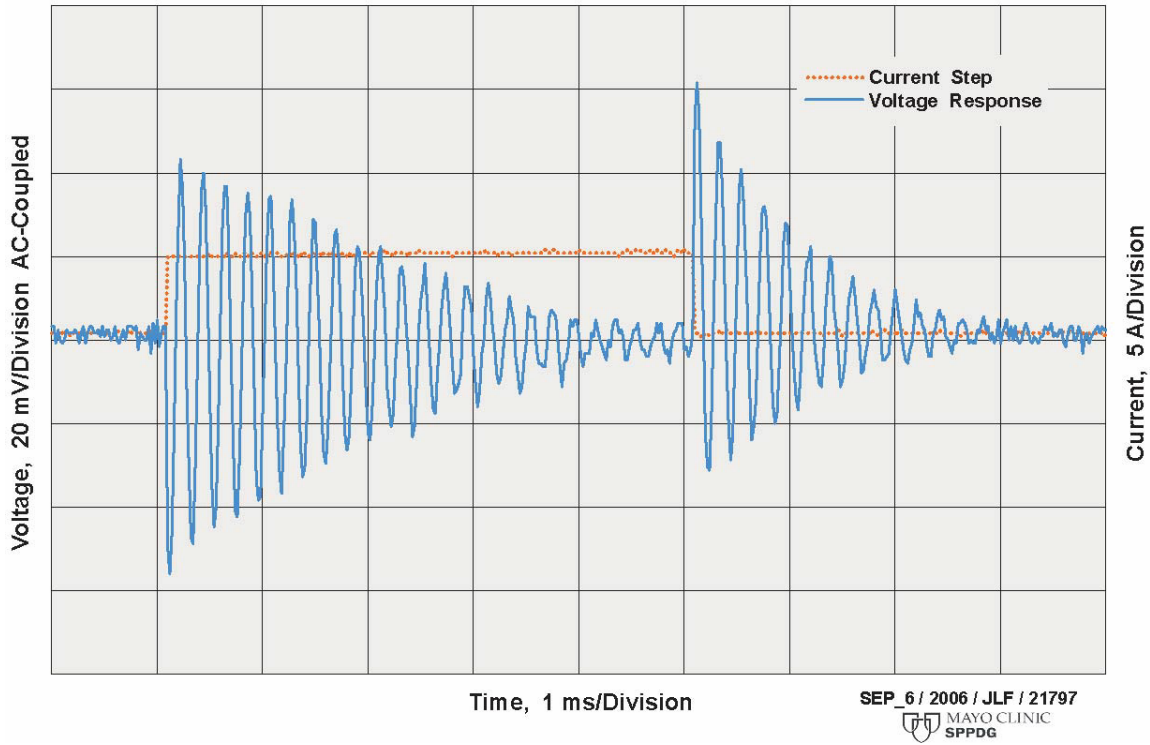

**Figure 1: Transient Response of Uncompensated Power Supply**

## Power Supply Stability Primer

A regulated power supply is actually a complete control system. While a thorough discussion of control theory is beyond the scope of this paper, it is necessary to introduce a few concepts. Control systems employ negative feedback techniques whereby the output of the system (Vout) is sampled (Vfb) and compared to the original input (Vref) as shown in Figure 2. The difference (Verror) stimulates the control system to correct itself. When the feed-forward path includes high gain, the magnitude of the error signal is driven towards zero and the output of the supply is made to track the reference. This tracking is largely independent from variations in the load current. It is possible to show [4] that the frequency-dependent closed-loop transfer function of the power supply is then given by

(1)    $V_{out}(s)/V_{ref}(s) = G(s)/[1+G(s)H(s)]$

where the feed forward response [$G(s)$] and feed back response [$H(s)$] are functions of frequency represented by the complex variable "s".

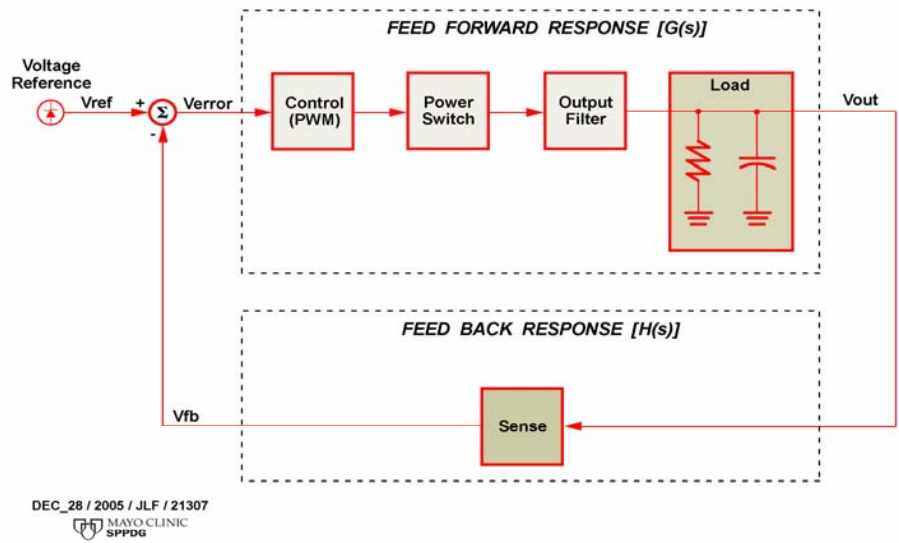

**Figure 2:  Voltage Regulator Block Diagram**

From equation (1) above, it is apparent that the quantity G(s)H(s) has a profound impact on the operation of the control system and from Figure 2, it is seen that this corresponds to the net gain around the open loop from Verror to Vfb.  This open loop gain or "loop gain" characterizes the dynamic performance of the supply.

Since the control system feeds its output back into its input, the possibility exists that certain values of loop gain may cause the output to become unstable.  In particular, when the magnitude of the loop gain is exactly equal to one (that is, 0 dB) and the phase shift of the loop gain is exactly 180 degrees, then the additional 180 degree phase shift from the inverting input of the summing node places the feedback signal in phase with the error signal and a sine wave oscillator is created.  Since self-oscillation is undesirable in a constant-voltage power supply, two metrics are defined to gauge how close a system may be to oscillation.  As seen in Figure 3, "Gain Margin" refers to the magnitude of the loop gain when the phase crosses the critical 180 degree threshold, while "Phase Margin" refers to the phase delay of the loop gain when the magnitude crosses the critical 0 dB threshold.  Well-designed systems typically have at least 10 dB of gain margin and at least 45 degrees of phase margin.

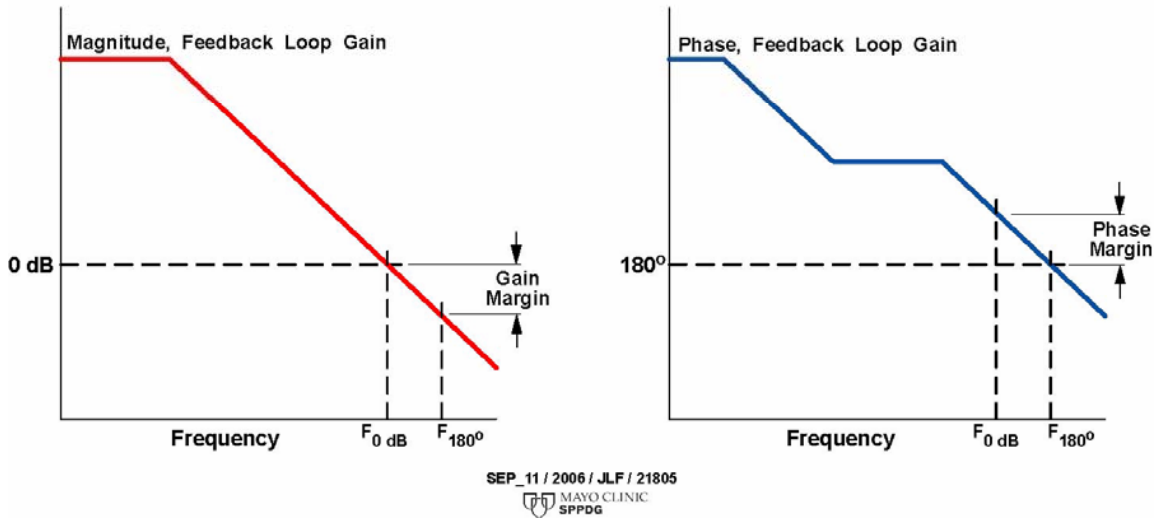

**Figure 3: Feedback System Stability Metrics**

Loop gain can be measured by adding a low-magnitude AC test signal to the feedback path [5], typically by inserting a transformer secondary winding in series between the load and sense circuit, as shown in Figure 4. This transformer is necessary because the feedback signal node (Vfb) is not at ground potential. Our analysis was conducted using a Tamura part number "MET-60" wideband audio transformer[1] as described in [11]. When applied to a "well-behaved" power supply, this measurement technique produces a magnitude and phase plot as shown in Figure 5. Using the definitions from Figure 3, we see that this supply has a gain margin of 28.6 dB and a phase margin of 90.2 degrees.

---

[1] Details regarding the transformer are curiously absent from [5], perhaps because the requirements depend upon the bandwidth of the control loop being investigated. Those interested in performing loop-gain measurements are encouraged to request documents [6] through [10] from their Agilent representative.

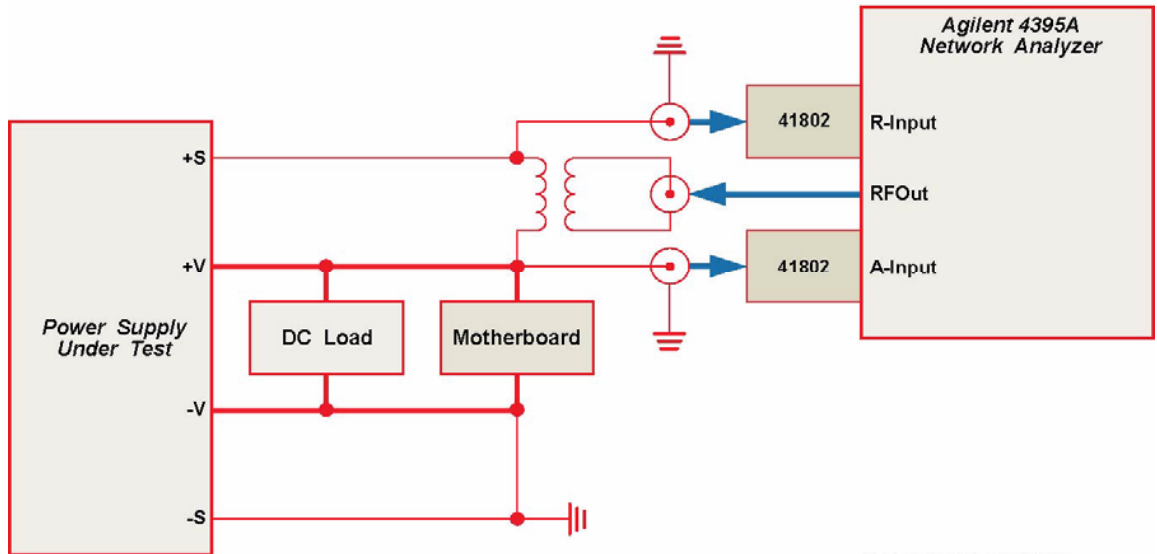

**Figure 4: Loop-Gain Measurement Setup**

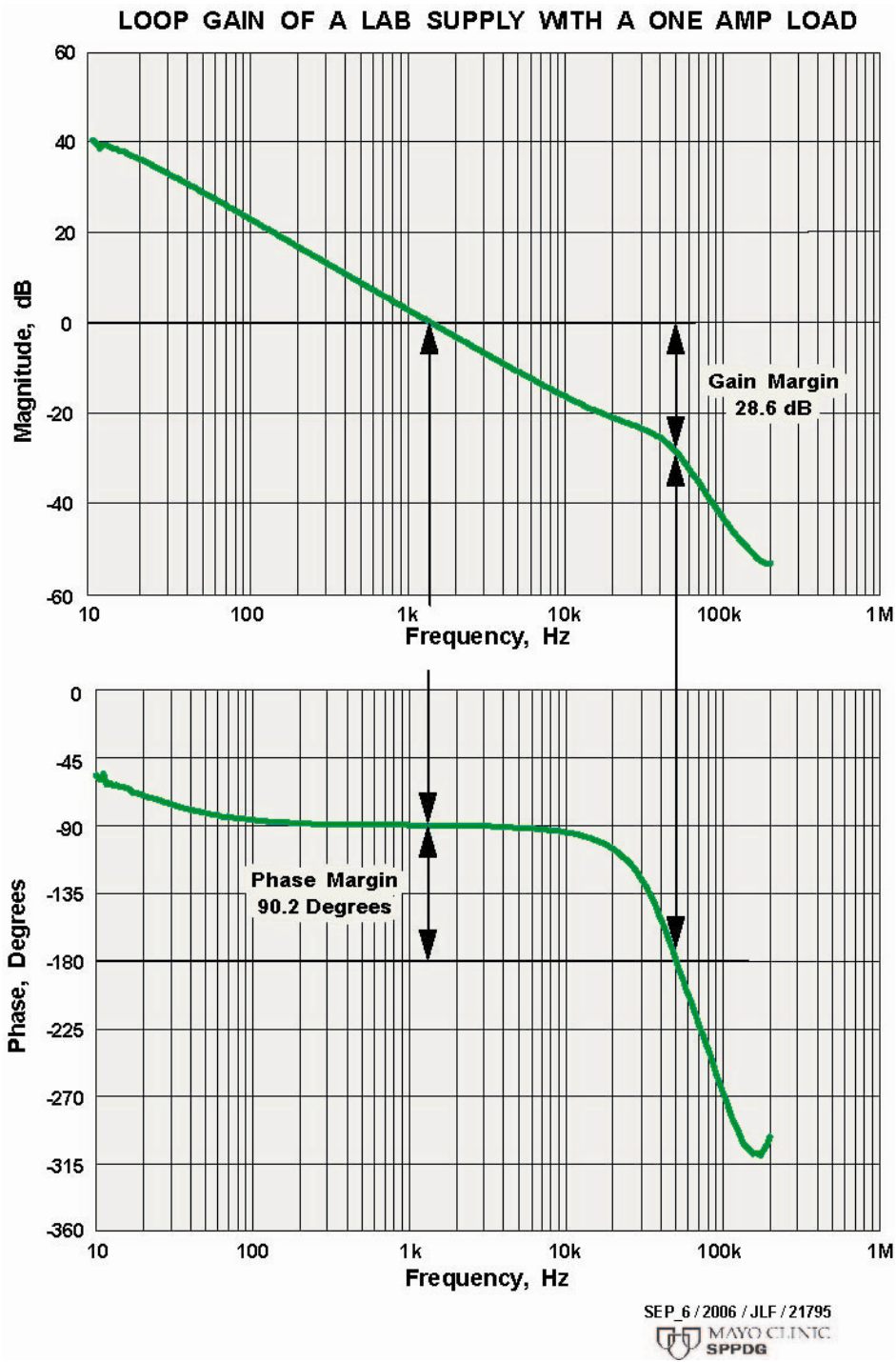

**Figure 5: Measured Loop Gain of a Well-Behaved Power Supply**

## Analysis

Insight into the internal supply stability issue may be obtained by referring to the magnitude and phase plots shown in Figure 6. From these plots, it is seen that the magnitude curve crosses 0 dB at 5.04 kHz, yielding only 13 degrees of phase margin, while the phase curve crosses -180 degrees at 6.05 kHz, yielding only 3.2 dB of gain margin.

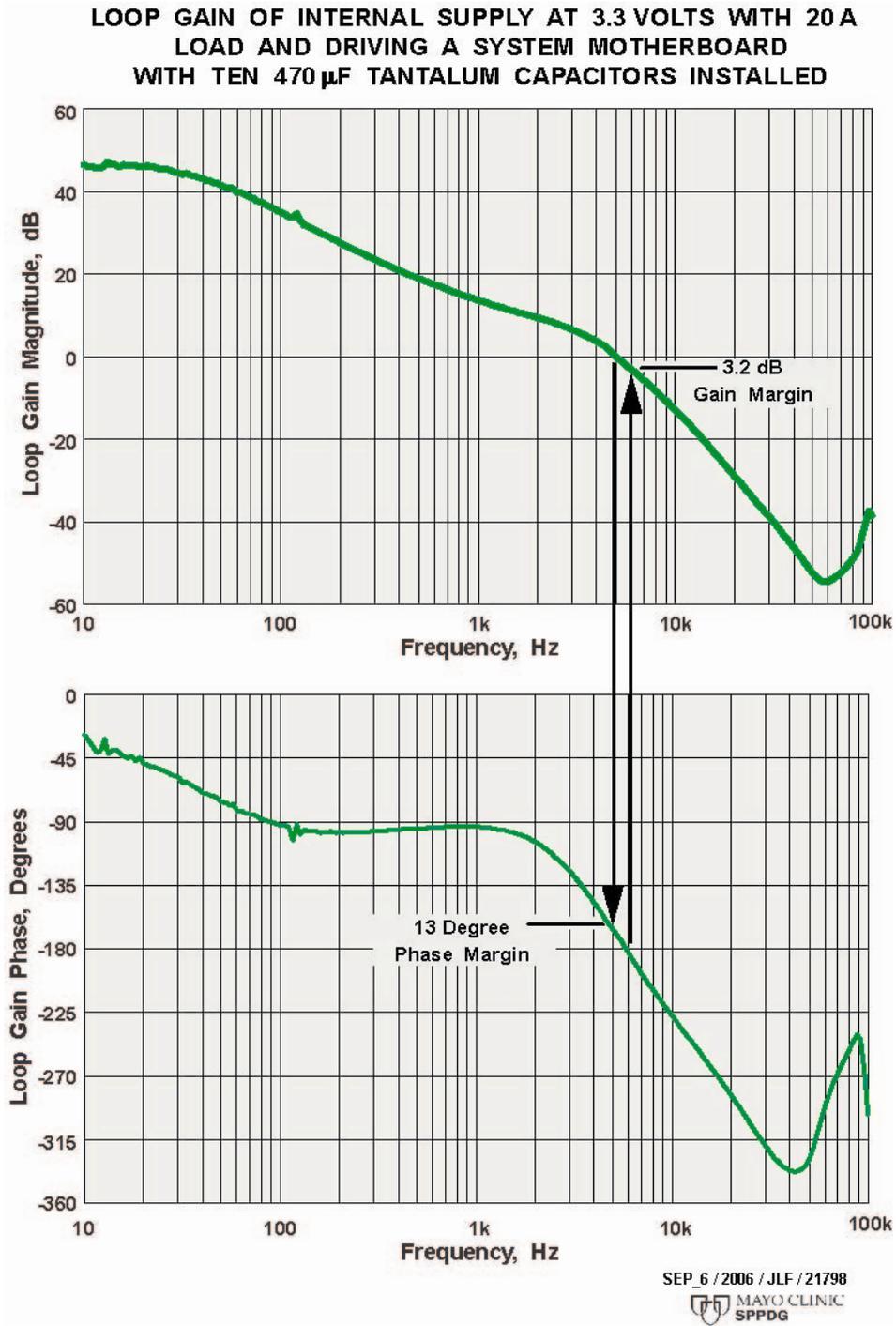

**Figure 6: Measured Loop Gain of Internal Supply**

The remedy chosen for this situation is the addition of a "phase-lead" network to the feedback path. A phase-lead network is essentially a high-pass filter that provides a low-impedance path from the supply output back to the sense (feedback) input without encountering the load. The effect of such a network is to postpone or delay the accumulation of phase until the loop-gain magnitude has passed safely through the 0 dB threshold. From the magnitude plot of the uncompensated loop we see that the curve increases its downward slope between 3 kHz and 4 kHz. For maximum effectiveness, the natural frequency of our phase-lead network should be at or below the 3 kHz value. For simplicity, we chose 2 kHz as our starting point.

Recalling that the natural frequency of a simple RC network is given by

(2)   $f_o = 1/2\pi RC$

and accounting for the fact that the power supply already has an internal 100 Ohm resistor (Rint) which parallels Rcomp in Figure 7, then it can be shown that

(3)   $R_{comp} = R_{int}/(2\pi f_o C_{comp} R_{int} - 1)$

Furthermore, we see that Rcomp is 20.38 Ohms when $f_o$ is 2 kHz and Ccomp is 4.7 microfarads. Close inspection of Figure 7 also reveals that we implemented the compensation network on the "-S" side of the feedback network so that the resistance of our injection transformer did not enter into the calculation of the compensation resistor value, and the frequency response of the supply did not change when the transformer was removed.

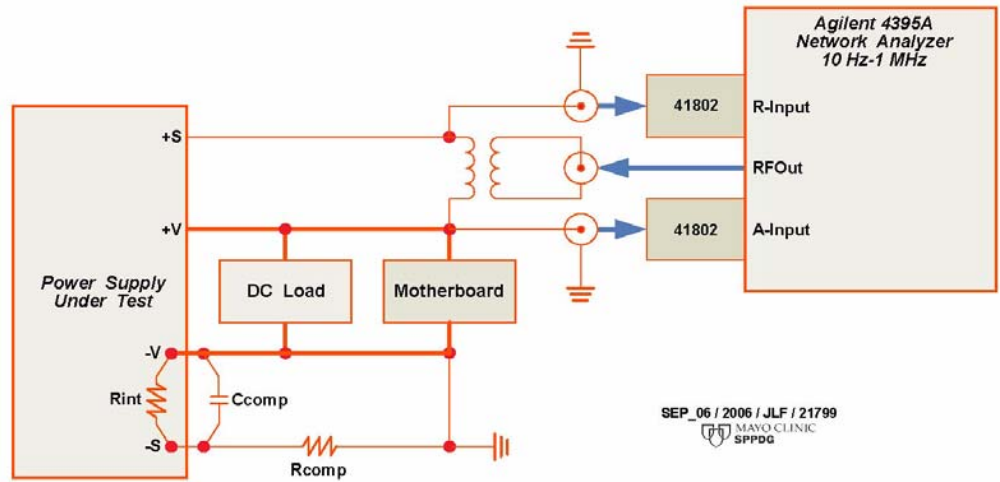

**Figure 7: Test Setup for Compensated Loop Gain Measurement**

The effect of this lead network is observed in the compensated loop-gain measurements shown in Figure 8. While the magnitude plot is not substantially affected, the -180 degree crossing of the phase plot is pushed out to 18.2 kHz (which is three times the uncompensated 0 dB crossover frequency), and the gain margin and phase margin are increased to 17 dB and 51 degrees respectively.

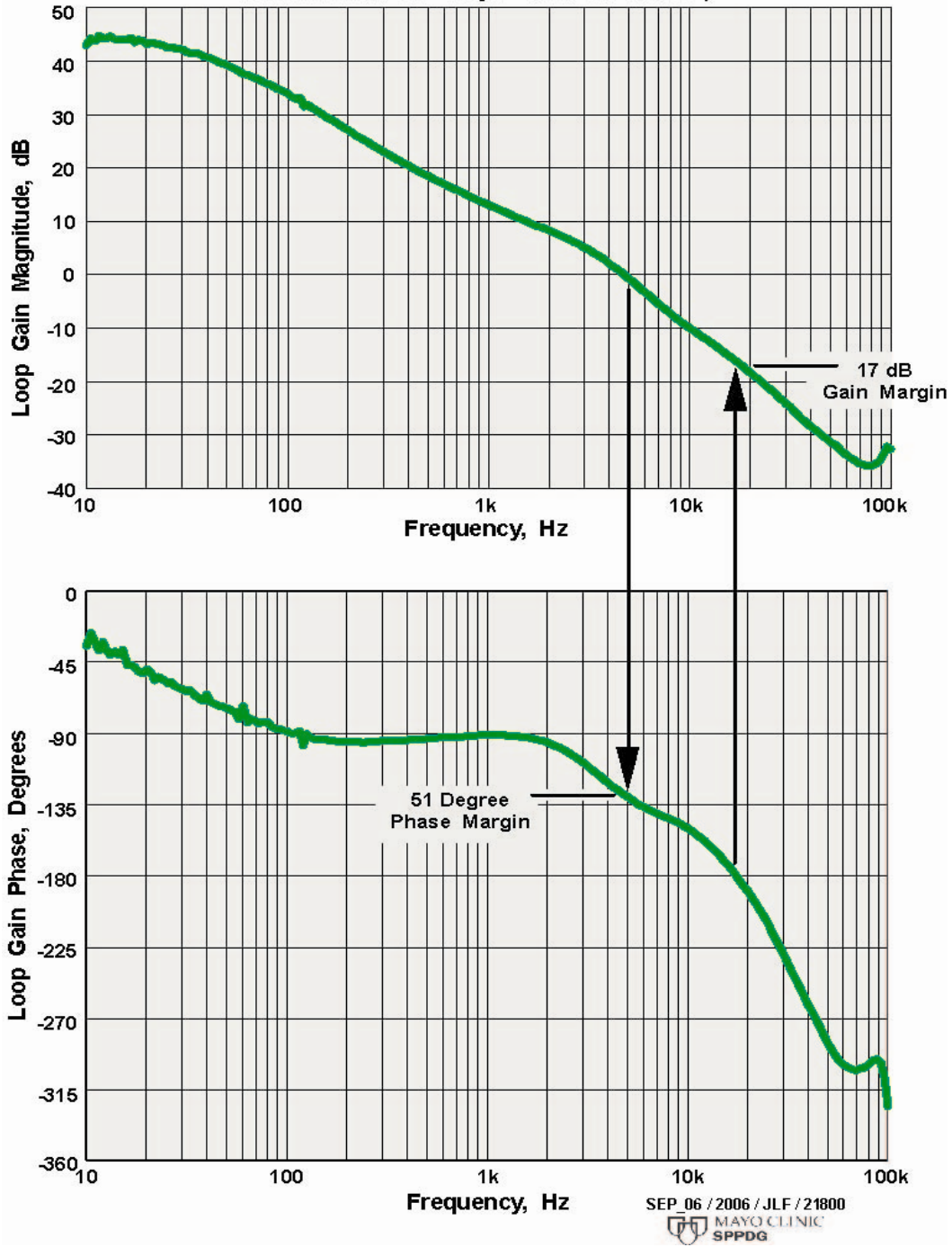

**Figure 8: Measured Loop Gain of Compensated Supply**

The effects of compensation are shown dramatically in the transient response shown in Figure 9.  The same 5 Ampere current step now produces a well-damped 60 mV peak (30 mV steady-state) droop in the supply voltage, while the persistent ringing of the supply voltage observed with the uncompensated supply has been eliminated.

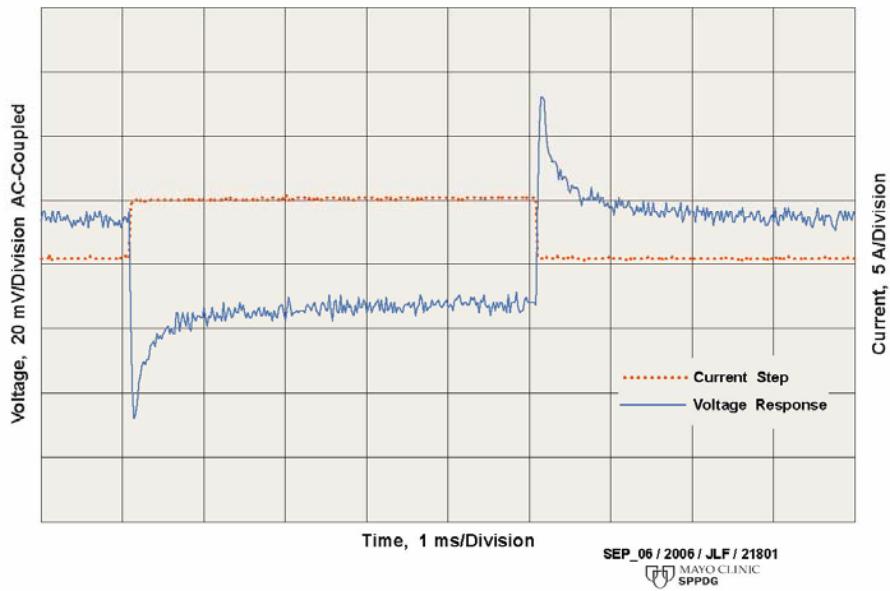

**Figure 9: Transient Response of the Compensated Supply**

## Conclusions

The primary lesson of this effort is the realization that decoupling design can no longer be done independently from power supply design or selection.  Traditional decoupling design strategy typically asserts that more capacitance is better, and from a narrow decoupling perspective that assertion is still valid.  However, the development of very-low ESR capacitors and fast-response regulators with wide loop-bandwidth means that it is now possible to accumulate loop phase so rapidly that regulator stability can be compromised.  In future systems the power distribution network and the power source must be developed concurrently and power system stability must be characterized.  Effective time-domain and frequency-domain techniques to facilitate this effort were presented.  In addition, a simple technique for power supply compensation was demonstrated.

## Acknowledgements

The authors would like to thank Elaine Doherty, Theresa Funk, Deanne Jensen, and Steve Richardson for their assistance in the creation of images for this report; and Jeffrey Bradley, Kevin Buchs, and Wendy Wilkins for their swift reviews of the manuscript.